\documentstyle[12pt]{article}
\begin{document}
\baselineskip=13pt
\begin{center} 
{\bf Effective Theory For Quantum Spin System In Low Dimension - Beyond
 Long-Wavelength Limit}
\end{center} 
\begin{center}
{\bf Ranjan Chaudhury}$^a$ and {\bf Samir K. Paul}$^b$\\
S. N. Bose National Centre For Basic Sciences\\
Block-JD, Sector-III, Salt Lake\\
Calcutta-700098,  India
\end{center}

\vspace{2.5cm}

\noindent        {\bf Abstract}

\hspace{0.2cm}

An effective theory for quantum spin system in low dimension is constructed

in the finite-q regime. It is shown that there are field configurations for 

which Wess-Zumino term contributes to the partition functions as topological

term for ferromagnet as well as antiferromagnet in both one and two

dimensional lattice,in contrast to the long wave length regime.


\vspace{1.5cm}

a)   {\bf {ranjan@boson.bose.res.in}}

b)   {\bf {smr@boson.bose.res.in}}

\newpage 

\vspace{0.5cm}
        
\noindent        {\bf Introduction}

\vspace{0.2cm}

       Quantum spin system in low dimension has drawn a lot of attention
 from the theoreticians after the discoveries of high temperature 
superconductivity  and quantum Hall effect.In a recent paper$^1$,we 
presented a unified scheme for analysing the topological terms in the effective
action corresponding to the long wave length limit for the XY-like anisotropic
quantum Heisenberg ferromagnet and antiferromagnet in one and two spatial
dimensions for any value of the spin.
\vspace{0.1cm}
         The present work originated from the apparent discrepancy between the
neutron scattering experimental results$^2$ and theoretical predictions
based on analysis$^3$ in the long wave length regime.Here we generalize
our previous formalism and analysis to medium wave length regime and 
characterize the excitations.The importance of the extension of the theoretical
analysis to the short wave length regime was also highlighted recently 
by Tai Kai Ng$^4$.
\vspace{0.1cm}
      Our results provide possible explanation of some of the qualitative
features observed in neutron scattering experiments.

\vspace{0.5cm}

\noindent       {\bf Mathematical Formulation}

\vspace{0.2cm}

  We analyse the quantum actions for XXZ ferromagnet and anti-ferromagnet in both
1D and 2D by 'spin coherent state' formalism$^{1,5}$ in the 'medium wave-length'
limit.This is the lowest order correction to the results obtained in the long
wave length limit.In particular we would like to investigate the existence
of topological term in the effective action corresponding to finite q-regime.As in
the previous paper$^1$ we choose the anisotropy of the spin models to be 
XY-like.

     We perform all the calculations on the lattice with a
finite lattice parameter $a$ in the so called 'medium wavelength' limit implying
that the Wess-Zumino term is kept up to order $a^2$ (in the long 
wave-length limit this was up to order a ) and the Hamiltonian is kept up to
order $a^3$ or $a^4$ according to whether the spin system is ferromagnetic or
antiferromagnetic(in the long-wavelength limit this was up to order 
$a^2$; for antiferromagnets the terms of the order $a^3$ turn out to be surface
 terms and hence vanish due to periodic boundary conditions ).
  
\vspace{0.5cm}

\noindent       {\bf Calculations}

\vspace{0.2cm}

The quantum Euclidean action ${S_E}[{\bf n}]$ for spin coherent fields ${\bf n}$ is given by$^{1,5}$ :
\begin{equation}
S_{E}[{\bf n}]=-is{\sum_{\bf r}}S_{WZ}[{\bf n}({\bf r})]+\frac{s\delta t}{4} {\sum_{\bf r}}\int^\beta_0 dt 
{({\partial_t}{\bf n}({\bf r}))^2} + \int^\beta_0 dt H({\bf n})
\end{equation}   
where s is the magnitude of the spin and
\begin{equation}
H({\bf n}) = \langle {\bf n} | H({\bf s}) |{\bf n}\rangle ; |{\bf n}\rangle ={{\prod}_{\otimes \bf r}} |{\bf n}({\bf r})\rangle  
\end{equation} 
$H({\bf s})$ being the spin Hamiltonian in the representation s.The Wess-Zumino
term  $S_{WZ}$ is given by$^5$
\begin{equation}
S_{WZ}[{\bf n}({\bf r})]={\int^\beta_0}dt{\int^1_0}{d\tau}{\bf n}({\bf r},t,{\tau})\cdot
{\partial_t}{\bf n}({\bf r},t,{\tau})\wedge{\partial_\tau}{\bf n}({\bf r},t,{\tau})={\mathcal{A}}
\end{equation} 
with ${\bf n}(t,0) \equiv {\bf n}(t)$,${\bf n}(t,1) \equiv \bf n_0$,
${\bf n}(0, \tau ) \equiv {\bf n}(\beta ,\tau )$; $t\in [0,\beta ]$,
$\tau\in [0,1]$. 

  In (3) $\mathcal{A}$ is the area of the cap bounded by the trajectory $\Gamma$
 parametrized by ${\bf n}(t)$ on the sphere

\begin{equation}
{\bf n}({\bf r})\cdot {\bf n}({\bf r})=1
\end{equation}

Here  $|{\bf n}\rangle$ is the spin coherent state as defined in Refs.1,5.The
spin Hamiltonian for XXZ Heisenberg ferromagnet (antiferromagnet)is given by

\begin{equation}
H({\bf S} ) = - g{\sum_{\langle {\bf r},{{\bf r}\prime} \rangle}} {\tilde
{\bf S}}({\bf r}) \cdot {\tilde{\bf S}}({{\bf r}\prime}) - g{\lambda_z}
{\sum_{\langle {\bf r},{{\bf r}\prime} \rangle}} {S_3}({\bf r}) {S_3}
({{\bf r}\prime})   
\end{equation}
with $g\ge 0$, (or $g\le 0$) $0\le \lambda_z \le 1$, where $\bf r$ and           ${{\bf r}\prime}$ run over
the lattice and $\langle {\bf r},{{\bf r}\prime} \rangle$ signifies nearest
neighbour interaction and ${\bf S} = (\tilde{\bf S}, S_3)$.

\vspace{0.5cm}

\noindent      {\bf Linear Chain}

\vspace{0.2cm}

Using eqn.(1) the quantum Euclidean action for ferromagnet in the\\               medium-wavelength limit  can be written as:      
\begin{eqnarray}
{S_E}[{\bf n}]& = & -is8{\pi}f[{\bf n}]\nonumber\\
&  & -isa{\sum_{i=1}^N}{\int^\beta_0}dt[{\bf m}\cdot {\partial_t}{\bf m}\wedge {\partial_x}{\bf m}](2ia)\nonumber\\ 
&  &-is{a^2}{\sum_{i=1}^N}{\int^\beta_0}dt[{\bf m}\cdot{\partial_t}{\bf l}\wedge
{\partial_x}{\bf m}\nonumber\\                                                  &  &+{\bf l}\cdot{\partial_t}{\bf m}\wedge{\partial_x}{\bf m}\nonumber\\
&  & +(-{\frac{1}{2}}{{\partial_x}^2}{\bf m}+{\partial_x}{\bf l})\cdot{\bf m}\wedge {\partial_t}{\bf m}](2ia)\nonumber\\                                        &  & +g{s^2}{a^2}{\sum_{i=1}^N}{\int^\beta_0}dt[{{({\partial_x}{\bf m})}^2}+2a{\partial_x}{\bf m}\cdot {\partial_x}{\bf l}](2ia)\nonumber\\
&  & - g{s^2}(1- {\lambda_z}){a^2}{\sum_{i=}^N}{\int^\beta_0} dt [{({\partial_x}
{\bf m})^2}+2a {\partial_x}{m_3}{\partial_x}{l_3}](2ia)\nonumber\\
&  &+ g{s^2}(1- {\lambda_z}){\sum_{i=1}^N}{\int^\beta_0} dt[2{{m_3}^2}+4a{l_3}{m_3}
+ 2{a^2}{{l_3}^2}](2ia)
\end{eqnarray}
where we have used the Hamiltonian in the spin coherent field($\bf n$) representation :
\begin{equation}
H({\bf n})=-g{s^2}{\sum_{i=1}^{2N}}[{\tilde{\bf n}}(ia)\cdot {\tilde{\bf n}}((i+1)a)-g{s^2}{\lambda_z}{\sum_{i=1}^{2N}}{n_3}(ia){n_3}((i+1)a)
\end{equation}
with ${\bf n}(ia)={\bf m}(ia)+a{\bf l}(ia)$,${\bf n}$ being the                 order parameter field and ${\bf l}$ the fluctuation arround ${\bf m}$. 
The first term in (6) viz.,$f[{\bf n}]$ is a fraction and depends on the ${\bf n}$-field configuration.However this being always a fraction it has no bearing on the winding number of a  topological sector.The second term on the right
hand side of (6)[denoted by $WZ$ in eqn.(8)] arises only from the Wess-Zumino
part of the action given by (1).This being a term in the order parameter field $\bf m$,the fractional term $f[{\bf n}]$ does not contribute to this term while integrating out the fluctuation field $\bf l$ from the partition function.We analytically continue action (6) to Minkowski one by replacing $t$ by $it$ and $\beta$ by $iT$.We take the truncated action without the fractional term and integrate out the field $\bf l$.The effective action in the order parameter field $\bf m$ takes the following form:

\begin{eqnarray}
{S_M}& = & \nonumber\\
&  &[sa WZ+s{a^2}{\sum_{i=1}^N}{\int^T_0}dt {\bigg( {{B_1}^2}+{{B_2}^2}+{\rho}{{C_3}^2} \bigg) } \nonumber\\
&  & -{\frac{s{a^2}}{2}}{\sum_{i=1}^N}{\int^T_0}dt{\bf m}\cdot {\partial_t}
{\bf m}\wedge {{\partial_x}^2}{\bf m}\nonumber\\
&  &-g{s^2}{a^2}{\sum_{i=1}^N}{\int^T_0}dt({{({\partial_x}{\tilde{\bf m}})}^2}+{\lambda_z}{{({\partial_x}{m_3})}^2})\nonumber\\
&  & -2g{s^2}(1-{\lambda_z}){\sum_{i=1}^N}{\int^T_0}dt{{m_3}^2}](2ia)
\end{eqnarray}

In eqn.(8) we have defined $WZ={\sum_{i=1}^N}{\int^T_0}dt({\bf m}\cdot {\partial_t}{\bf m}\wedge {\partial_x}{\bf m})$ ;\\                                      ${B_k}=-{{{\frac{1}{a}}{\bf m}+{\frac{3}{2}}{\partial_t}{\bf m}\wedge {\partial_x}{\bf m}+gsa{{\partial_x}^2}{\bf m}}_k}$\\
 ; $k=1,2$ ;${C_3}={{-{\frac{1}{a}}{\bf m}+{\frac{3}{2{\rho}}}{\partial_t}{\bf m}\wedge {\partial_x}{\bf m}+{\frac{gsa{\lambda_z}}{\rho}}{{\partial_x}^2}{\bf m}}_3}$ and ${\rho}=1+2gs(1-{\lambda_z})$.\\
Now we analytically continue the expression on right hand side of eqn.(8) to Euclidean space and demand finiteness of the action.As a consequence the boundary  
conditions on the ${\bf m}$-fields are quite general.However the class of field configurations satisfying the boundary conditions  ${{\bf m}(x,t)}\longrightarrow {{\bf m}_0}$ at $\infty$ [Ref.1] together with ${m_3}\longrightarrow 0$ at $\infty$ keeps the action finite and makes the $WZ$ term in eqn.(8) a topological one$^{1,5}$. 
\noindent  The quantum action for antiferromagnet in the medium wavelength limit can be written as :
\begin{eqnarray}                                                                
{S_E}[{\bf n}] & = & - isa{\sum_{i=1}^N}{\int^\beta_0} dt \nonumber\\           &  &[{\bf m}\cdot {\partial_t}{\bf m}\wedge {\partial_x}{\bf m}+2{\bf l}\cdot  
{\bf m}\wedge {\partial_t}{\bf m}](2ia) \nonumber\\                             &  &-is{a^2}{\sum_{i=1}^N}{\int^\beta_0}dt [ - {\frac{1}{2}}{\bf m} 
\cdot {\partial_t}{\bf m}\wedge {{\partial_x}^2}{\bf m}\nonumber\\
&  & - {\bf m}\cdot {\partial_t}{\bf m}\wedge {\partial_x}{\bf l} 
- {\bf m}\cdot {\partial_x}{\bf m}\wedge {\partial_t}{\bf l} \nonumber\\         &  & + 2{\bf l}\cdot {\bf m}\wedge {\partial_t}{\bf l} + {\bf l}\cdot           {\partial_t}{\bf m}\wedge {\partial_x}{\bf m}](2ia)\nonumber\\
&  &+g{s^2}{a^2}{\sum_{i=1}^N}{\int^\beta_0}dt[4{{\bf l}^2} + ({{{\partial_x}{\bf m}})^2}\nonumber\\
&  & - {\frac{a^2}{12}}{({{{\partial_x}^2}\bf m})^2} + {a^2}{\bf l}\cdot
{{\partial_x}^2}{\bf l}](2ia)-g{s^2}(1-{\lambda_z}){a^2}{\sum_{i=1}^N}\nonumber\\
&  &{\int^\beta_0}dt[4{{l_3}^2}+{{{\partial_x}{m_3}}^2}-{\frac{a^2}{12}}
{({{{\partial_x}^2}{m_3}})^2} + {a^2}{l_3}{{{\partial_x}^2}{l_3}}](2ia)\nonumber\\   
&  & + 2g{s^2}(1-{\lambda_z}){\sum_{i=1}^N}{\int^\beta_0} dt ({{m_3}^2} + {a^2}
{{l_3}^2})(2ia)
\end{eqnarray}      

where we have used in eqn.(1) 
\begin{equation}
H({\bf n}) = g{s^2}{\sum_{i=1}^{2N}} {\tilde {\bf n}}(ia)\cdot {\tilde {\bf n}}(
(i+1)a) + g{s^2}{\lambda_z}{\sum_{i=1}^{2N}}{n_3}(ia){n_3}((i+1)a)
\end{equation}

The expression of the above action (9) is different from (6) due to the fact 
that in the Hamiltonian (10) as well as in the Wess-Zumino term ${S_{WZ}}
[{\bf n}(ia)]$ we have done the staggering operation:
${\bf n}(ia)\longrightarrow {(-1)^i}{\bf n}(ia)$ and the staggered field ${\bf n}(ia)$ is then set ${\bf n}(ia) = {\bf m}(ia) + {(-1)^i}a {\bf l}(ia)$.We neglect terms like ${\bf l}\cdot {{\partial_x}^2}{\bf l}$ in the above action since $\bf m$ is slowly varying and $\bf l$ behaves as a derivative of $\bf m$ and as a result these terms are smaller with respect to terms involving derivatives of $\bf m$ within the terms of the order ${a^4}$.\\
Then we continue the action (9) to Minkowski one in a similar fashion as in the case of ferromagnet and integrate out the field $\bf l$ from the partition function.The effective action in the order parameter field takes the following form :
\begin{eqnarray}
{S_M}& = &[sa WZ +{\frac{1}{16g}}{\sum_{i=1}^N}{\int^T_0}dt\bigg( {{D_1}^2}+{{D_2}^2}+{\frac{2}{(1+{\lambda})}}{{D^3}^2} \bigg) \nonumber\\ 
&  &-g{s^2}{a^2}{\sum_{i=1}^N}{\int^T_0}dt \bigg( {{E_1}^2}+{{E_2}^2}+{\lambda_z}{{E_3}^2} \bigg) \nonumber\\
&  &-s{a^2}{\sum_{i=1}^N}{\int^T_0}dt{\frac{1}{2}}{\bf m}\cdot {\partial_t}       {\bf m}\wedge {{\partial_x}^2}{\bf m}\nonumber\\
&  &-2g{s^2}{(1-{\lambda_z})}{\sum_{i=1}^N}{\int^T_0}dt{{m_3}^2}](2ia)
\end{eqnarray}
where ${\bf D}=2{\bf m}\wedge {\partial_t}{\bf m}+a{\bf m}\wedge {\partial_t}{\partial_x}{\bf m}$,$k=1,2$\\
${E_l}=({\partial_x}{m_l})({\partial_x}{m_l})-{\frac{a^2}{12}}({{\partial_x}^2}{m_l})({{\partial_x}^2}{m_l})$,$l=1,2,3$.\\

Again the argument similar to the one in the case of ferromagnet, the field configurations satisfying ${\bf m}\longrightarrow {{\bf m}_0}$ at $\infty$ with ${m_3}\longrightarrow 0$ at $\infty$ makes the $WZ$ term
 in eqn.(11) a topological one.\\ 
    At this stage we point out that in the long wavelength limit i.e,when we 
keep terms up to order $a$ in the WZ-part and up to order $a^2$ in the Hamiltonian part 
 and integrate out the fluctuation ${\bf l}$ in the partition
function we get usual
effective action in $\bf m$ .This leads to non-linear sigma model with a topological term in the case of
antiferromagnet and a different model with a topological term in the case of ferromagnet . 

\vspace{0.5cm}

\noindent    {\bf Two dimensional square lattice}

\vspace{0.2cm}

   The Minkowskian action for ferromagnet in the two dimentional square lattice
can be written in the form:

\begin{eqnarray}
{S_M}& = & 8{\pi}s{\chi}[{\bf n}] + sa{\sum_{i,j=1}^N}{\int^T_0}dt[{\bf m}\cdot
{\partial_t}{\bf m}\wedge {\partial_x}{\bf m}\nonumber\\
&   &{\bf m}\cdot {\partial_t}{\bf m}\wedge {\partial_x}{\bf m}](2ia,2ja) - s{a^2}
{\sum_{i,j=1}^N}{\int^T_0}dt[{{\bf l}^2} + 2{\bf l}\cdot \nonumber\\
&   &{ {\frac {\bf m}{a}} - {\frac {3}{2}}{\bf M} } + {\frac {1}{2}}{\bf m}
\cdot {\partial_t}{\bf m}\wedge {{\bigtriangledown }^2}{\bf m}](2ia,2ja)\nonumber\\
&   &-2g{s^2}{a^2}{\sum_{i,j=1}^N}{\int^T_0}dt[{({\partial_x}{m_1})^2}+
{({\partial_x}{m_2})^2} + {({\partial_y}{m_1})^2} \nonumber\\
&   &{({\partial_y}{m_2})^2} - 2a {\bf {\tilde l}}\cdot {{\bigtriangledown }^2}
{\bf {\tilde m}}](2ia,2ja) - 2g{s^2}{\lambda_z}{a^2}{\sum_{i,j=1}^N}\nonumber\\
&   &{\int^T_0}dt[{({\partial_x}{m_3})^2} + {({\partial_y}{m_3})^2}-2a{l_3}
{{\bigtriangledown }^2} {m_3}](2ia,2ja) - 2g{s^2}(1-{\lambda_z})   \nonumber\\
&   &{\sum_{i,j=1}^N}{\int^T_0}dt[4{{m_3}^2} + 8a{m_3}{l_3} + 4{a^2}{{l_3}^2}](2ia,2ja)
\end{eqnarray}
Here ${\chi}[{\bf n}]$ is a proper fraction depending on the ${\bf n}$ field
configuration.${\bf M}={\partial_t}{\bf m}\wedge ({\partial_x}{\bf m}+{\partial_y}{\bf m})$ and $m_i$,$l_i$,$M_i$ stand for the  $i$th components of ${\bf m}$,${\bf l}$,${\bf M}$ in the spin space.${\tilde {\bf m}}$,${\tilde {\bf l}}$,${\tilde {\bf M}}$ stand for the vectors  in the XY-plane of spin space.\\      
Using the same argument as used in the case of ferromagnetic chain we neglect the fractional number ${\chi}[{\bf n}]$ and work with the reduced action.After integrating out the field $\bf l$, the effective action in the order parameter field $\bf m$ takes the form:

\begin{eqnarray}
{S_M}& = & sa WZ +s{a^2}{\sum_{i,j=1}^N}{\int^T_0}dt[{{T_1}^2}+{{T_2}^2}+\alpha {{R_3}^2}](2ia,2ja)\nonumber\\
&  &-{\frac{s{a^2}}{2}}{\sum_{i.j=1}^N}{\int^T_0}dt[{\bf m}\cdot {\partial_t}{\bf m}\wedge {{\bigtriangledown}^2}{\bf m}](2ia,2ja)\nonumber\\
&  &-2g{s^2}{a^2}{\sum_{i,j=1}^N}{\int^T_0}dt[{({\partial_x}{\tilde{\bf m}})^2}+{({\partial_y}{\tilde{\bf m}})^2}\nonumber\\
&  & +{\lambda_z}{({\partial_x}{m_3})^2}+{\lambda_z}{({\partial_y}{m_3})^2}](2ia,2ja)\nonumber\\
&  &-{\frac{1}{2}}s{a^2}{\sum_{i,j=1}^N}{\int^T_0}dt ({\bf m}\cdot {\partial_t}
{\bf m}\wedge {{\bigtriangledown}^2}{\bf m})(2ia,2ja)\nonumber\\
&  & -8g{s^2}(1-{\lambda_z}){\sum_{i,j=1}^N}{\int^T_0}dt{{m_3}^2}(2ia,2ja)
\end{eqnarray}
where $WZ=[{\bf m}\cdot {\partial_t}{\bf m}\wedge {\partial_x}{\bf m}
+{\bf m}\cdot {\partial_t}{\bf m}\wedge {\partial_y}{\bf m}](2ia,2ja)$,\\
${T_k}={\frac{1}{a}}{m_k}-{\frac{3}{2}}{M_k}-2gsa{{\bigtriangledown}^2}{m_k}$,
$k=1,2$\\
and ${R_3}={\frac{1}{a}}{m_3}-{\frac{3}{2\alpha}}{M_3}-{\frac{2gsa{\lambda_z}}{\alpha}}
{{\bigtriangledown}^2}{m_3}$
Here ${{\bigtriangledown}^2}$ stands for ${{\partial_x}^2}+{{\partial_y}^2}$ in two dimensional space and $\alpha =1+8gs(1-{\lambda_z})$.\\ 
We  write down the action for the antiferromagnet neglecting the terms
${\partial_x}{{\bf l}^2}$,${\partial_y}{{\bf l}^2}$ as compared to terms involving $\bf m$ or derivatives of $\bf m$ within $a^4$ order in a manner similar to the case of antiferromagnetic chain as:

\begin{eqnarray}
{S_M}& = & s{\sum_{i,j=1}^N}{\int^T_0}dt[4a{\bf l}\cdot {\bf m}\wedge
{\partial_t}{\bf m}](2ia,2ja)+s{a^2}{\sum_{i,j=1}^N}{\int^T_0}dt[-2{\bf l}\cdot {\partial_t}
\nonumber\\ 
&  &{\partial_x}{\bf m}\wedge {\bf m} - 2{\bf l}\cdot {\partial_t}\nonumber\\   &  & \wedge {\partial_y}{\bf m} + {\bf l}\cdot {\partial_t}{\bf m}\wedge 
{\partial_x}{\bf m} + {\bf l}\cdot {\partial_t}{\bf m}\wedge \nonumber\\        &  & {\partial_y}{\bf m}](2ia,2ja)+s{\frac{1}{2}}{a^2}{\sum_{i,j=1}^N}{\int^T_0}dt[{\partial_x}
({\bf m}\cdot{\partial_t}{\bf m}\wedge {\partial_y}({\bf m}\cdot{\partial_t}
{\bf m}\wedge {\partial_x}{\bf m})](2ia,2ja)\nonumber\\
&  & - 2g{s^2}{a^2}{\sum_{i,j=1}^N}{\int^T_0}dt[8{{\bf l}^2}+ {({\partial_x}{\bf m})^2} 
+{({\partial_y}{\bf m})^2}](2ia,2ja)-2g{s^2}{a^2}(1-{\lambda_z}){\sum_{i,j=1}^N}\nonumber\\
&  &{\int^T_0}dt[8{{l_3}^2} + {({\partial_x}{m_3})^2} + {({\partial_y}{m_3})^2}](2ia,2ja)\nonumber\\
&  & - g{s^2}{a^4}{\sum_{i,j=1}^N}{\int^T_0}dt[{\frac{1}{6}}{({{\partial_x}^2}{\bf m})^2}
+ \frac{1}{6} ({\partial_y^2}{\bf m})^2](2ia,2ja)\nonumber\\
&  & + g{s^2}{a^4}(1-{\lambda_z}){\sum_{i,j=1}^N}{\int^T_0}dt[\frac{1}{6} {({\partial_x^2} 
{m_3})^2}](2ia,2ja)    
\end{eqnarray}
After integrating out the fluctuation field ${\bf l}$ we have the following
form of the effective action:
\begin{eqnarray}
{S_M}& = &[{\frac{s{a^2}}{2}} WZ+{\frac{1}{4g}}{\sum_{i,j=1}^N}{\int^T_0}dt({{K_1}^2}+{{K_2}^2}+{\frac{4}{1+3{\lambda_z}}}{{K_3}^2})\nonumber\\ 
&  &-2g{s^2}{a^2}{\sum_{i,j=1}^N}{\int^T_0}dt({{J_1}^2}+{{J_2}^2}+{\lambda_z}{{J_3}^2})\nonumber\\
&  &+{\frac{g{s^2}{a^4}}{6}}{\sum_{i,j=1}^N}{\int^T_0}dt({{P_1}^2}+{{P_2}^2}+{\lambda_z}{{P_3}^2})\nonumber\\
&  &-{\frac{g{s^2}{a^4}(1-{\lambda_z})}{12}}{\sum_{i,j=1}^N}{\int^T_0}dt({{P_3}^2}+
+12{({\partial_x}{\partial_y}{m_3})^2})\nonumber\\ 
&  &-4g{s^2}(1-{\lambda_z}){\sum_{i,j=1}^N}{\int^T_0}dt{{m_3}^2}](2ia,2ja) 
\end{eqnarray}
Where $WZ={\sum_{i,j=1}^N}{\int^T_0}dt[{\partial_x}({\bf m}\cdot {\partial_t}{\bf m}\wedge {\partial_y}{\bf m})
+{\partial_y}({\bf m}\cdot {\partial_t}{\bf m}\wedge {\partial_x}{\bf m})]$,\\
${\bf K}=[{\bf m}\wedge {\partial_t}{\bf m}+{\frac{a}{2}}({\bf m}
\wedge {\partial_t}{\partial_x}{\bf m}+{\bf m}\wedge {\partial_t}{\partial_y}
{\bf m})+{\frac{a}{4}}{\bf M}]$,\\                                              ${\bf M}$ has been defined before in the action for ferromagnet.\\
${{J_K}^2}={({{\partial_x}{m_k}})^2}+{({{\partial_y}{m_k}})^2}$,
${{P_k}^2}={({{{\partial_x}^2}{m_k}})^2}+{({{{\partial_y}^2}{m_k}})^2}$ \\
with $k=1,2,3$

To see the role of WZ-term in ferromagnet as well as antiferromagnet we go back to Euclidean forms of the actions (13) and (15) and demand finiteness of the actions.As before the class of $\bf m$-field configurations satisfying boundary conditions ${\bf m}\longrightarrow {{\bf m}_0}$ at $\infty$ and ${m_3}\longrightarrow 0$ at $\infty$ makes the WZ-term topological in both the cases.\\
The periodic boundary condition on the space lattice and  time makes ${\bf m}$ periodic but not necessarily their derivatives.Thus the  WZ-term on the right hand side of eqn.(15) is a topological term as it is derivative of the winding number which has been dicussed in Ref.1.

\vspace{1.0cm}

\noindent    {\bf Conclusions and comments}
 
\vspace{0.5cm}

The lattice structure has to be explicitly dealt with as we are working with
finite-q (in the medium and short wave length regime).The definition of
topological term  is however exact  only when we go to continuum limit$(q\longrightarrow 0)$ .         

\vspace{0.1cm}

There should be a cross-over from non-topological to topological behaviour
in physical manifestation for 2D quantum Heisenberg antiferromagnet as we increase the magnitude of the wave vector (i.e., q-value) to go from long wave length  regime to short wave length regime.The neutron scattering experiments performed on the Copper Oxides in the antiferromagnetic phase,seem to probe the medium and the short wave-length regime and exihibit some of the features of topological excitations $^{1,2,5}$ in accordance with our inference.   

\vspace{0.1cm}

   The quantum Kosterlitz-Thouless-like scenerio may hold in the finite-q regime too,as the topological excitations continue to exist in this regime for a 2D quantum Heisenberg ferromagnet$^1$.  

\noindent          {\bf References}
\vspace{0.2cm}

\noindent 1

Ranjan Chaudhury and Samir K. Paul, Phys.Rev.B60,6234(1999)

\vspace{0.5cm}

\noindent 2

Y.Endoh et al,Phys.Rev.B37,7443(1988);   K.Yamada et al.,ibid.40,               4557(1989);\\                                                                   R.Chaudhury,Indian J.Phys.,66A,159(1992)

\vspace{0.5cm}

\noindent 3

S.Chakravarty,B.I.Halperin and D.R.Nelson,Phys.Rev.Lett.60,1057(1988);\\        S.Tyc,B.I.Halperin and S.Chakravarty,ibid.62,835(1989);\\                       A.Auerbach and D.P.Arovas,ibid.61,617(1988)

\vspace{0.5cm}

\noindent 4

Tai Kai Ng.,Phys.Rev.Lett.,82,3504(1999)

\vspace{0.5cm}

\noindent 5

E.Fradkin and M.Stone,Phys.Rev.B38,7215(1988);\\
E.Fradkin,Field Theories of Condensed Matter Systems(Addison-Wesly,CA,1991) 

\end{document}